# Performance Evaluation of Ballistic Silicon Nanowire Transistors with Atomic-basis Dispersion Relations

Jing Wang[*], Anisur Rahman, Avik Ghosh, Gerhard Klimeck and Mark Lundstrom
School of Electrical and Computer Engineering, Purdue University, West Lafayette, Indiana 47907, USA. Email[*]: jingw@purdue.edu

## ABSTRACT

In this letter, we explore the bandstructure effects on the performance of ballistic silicon nanowire transistors (SNWTs). The energy dispersion relations for silicon nanowires are evaluated with an $sp^3d^5s^*$ tight binding model. Based on the calculated dispersion relations, the ballistic currents for both n-type and p-type SNWTs are evaluated by using a semi-numerical ballistic model. For large diameter nanowires, we find that the ballistic p-SNWT delivers half the ON-current of a ballistic n-SNWT. For small diameters, however, the ON-current of the p-type SNWT approaches that of its n-type counterpart. Finally, the carrier injection velocity for SNWTs is compared with those for planar metal-oxide-semiconductor field-effect transistors, clearly demonstrating the impact of quantum confinement on the performance limits of SNWTs.

With the rapid progress in nanofabrication technology, silicon nanowires (SiNWs) with small diameters (<20nm) have been synthesized and extensively studied for potential applications in nanoelectronics.[1-5] Among them, the silicon nanowire transistor (SNWT) has attracted broad attention as a promising structure for future integrated circuits due to its excellent scaling capability and compatibility with Si-based electronic technology.[1, 5-7] Device theory stipulates that to maintain a good electrostatic gate control, the diameter of a SNWT should be reduced as the gate length of the transistor scales down. For this reason, SNWTs with ultra-small diameters (<5nm) are needed when the gate lengths enter the sub-10nm regime.[7-8] Due to strong quantum confinement (QC) in ultra-thin SiNWs, atomic bandstructure effects[9-11] are expected to play an important role on their device characteristics.

In this letter, we theoretically explore the impact of bandstructure, nanowire diameter and carrier type (n- or p-type) on the performance limits of SNWTs. First, the energy dispersion (E-k) relations for SiNWs with different wire widths (e.g., 0.5-6.8nm) are calculated by using an $sp^3d^5s^*$ semi-empirical tight-binding approach.[10-12] A semi-numerical ballistic model[13] is then employed to evaluate the ballistic current-voltage (*I-V*) characteristics of both n-type and p-type SNWTs based on the calculated E-k relations. Finally, the carrier injection velocity for the simulated SNWTs is compared with that for the corresponding metal-oxide-semiconductor field-effect transistors (MOSFETs).

Silicon nanowires with various cross-section shapes and transport orientations have been synthesized by different experimental groups.[1-5] In this work, we focus on one specific SiNW structure with one particular transport orientation as a first step in exploring bandstructure effects in small SNWTs. The inset of Fig. 1 shows the cross section of the simulated SiNWs. The shape of the cross section is rectangular, transport is along the [100] direction, and the faces of the rectangle are along the equivalent <100> axes. The Si body thickness, $T_{Si}$, is assumed to be equal to the wire width, $W_{Si}$ (i.e., $T_{Si}= W_{Si}=D$). A unit cell of the SiNW crystal consists of four atomic layers along the *x* (transport) direction and has a length of $a_0$=5.43Å. In the nanowire Hamiltonian, each atom is modeled using 20 orbitals – the $sp^3d^5s^*$ basis with the spin-orbital coupling. The orbital-coupling parameters used in this work are from Ref. 12, which have been optimized by Boykin *et al.* to accurately reproduce bulk Si properties (band gap, effective-masses, *etc*). A hard wall boundary condition for the wavefunction is used at the semiconductor-oxide interfaces and the dangling bonds at these interfaces are pacified using a hydrogen-like termination model of the sp$^3$ hybridized interface atoms.[14] As demonstrated in Ref. 14, this technique successfully removes all the interface states from the band gap.

Figure 1 shows the E-k dispersion relations for the simulated SiNWs with (a) *D*=1.36nm and (b) *D*=5.15nm. The bulk conduction band of Si has six equivalent Δ valleys located near the X points in the Brillouin zone. For a nanowire with the transport axis along [100], four of the six equivalent Δ valleys (i.e., [010], [0$\bar{1}$0], [001] and [00$\bar{1}$]) are projected into the Γ point in the one dimensional Brillouin zone to form the conduction band edge. The other two Δ valleys (i.e., [100] and [$\bar{1}$00]), located at $k_x = \pm 0.815 \cdot 2\pi/a_0 = \pm 1.63\pi/a_0$ in the bulk Brillouin zone, are zone-folded to the points $k_x = \pm 0.37\pi/a_0$ in the wire Brillouin zone and become off-Γ states. A similar phenomenon is observed by Ko *et al.*[10] and Zheng *et al.*[11] in a [100] oriented rectangular wire with equivalent <110> confinement directions. As in a Si quantum well, the degeneracy of the 4-fold Γ valleys in a [100] oriented nanowire can be lifted by the interaction between the four equivalent valleys, which is so called "band splitting".[15] It is clear from Fig. 1 that the corresponding band splitting is more evident in the thinner wire (*D*=1.36nm) than in the thicker wire (*D*=5.15nm), analogous to the band splitting observed in Si quantum wells.[15]

Figure 2 shows the conduction and valence band edges (solid with circles) for the simulated wires with a wire width ranging from 1.0nm to 6.8nm. It shows that the wire band gap is enlarged by QC and the increment is *roughly* inversely proportional to the square of the wire width.[10-11] The dashed line in the upper plot is for the conduction band edge calculated by a simple "particle in a box" model with the bulk effective masses. The difference between the effective-mass (EM) curve and the tight binding result becomes evident when *D*<3nm, which shows that the bandstructure effects can be important in Si nanowires with small diameters.[9-11]

After obtaining the tight binding E-k relations of the simulated SiNWs, the *I-V* characteristics of the corresponding SNWTs can be explored by a semi-numerical ballistic model.[13] The model captures three-dimensional electrostatics, quantum capacitance[16] and bias-charge self-consistency in ballistic field-effect transistors (FETs). (Source-to-drain tunneling is not considered in this model). In the past, this model was used to evaluate the *I-V* characteristics of Si MOSFETs[13] and high electron mobility transistors[17] with parabolic energy bands and Ge MOSFETs with numerical E-k relations[18]. To be concise, we do not indicate the details of this model but refer the readers to published references.[13, 17, 18] (The Matlab® scripts of this model are available.[19])

To compare the device performance of n-type *vs.* p-type SNWTs, we adjust the gate work functions to achieve the same OFF-currents for the two structures at each wire width (see Fig. 3 (a) for an example of *D*=1.36nm), then the ON-currents for both the nFET and pFET are compared. The inset of Fig. 3 (a) plots the ratio of the pFET ON-current to that for the nFET at the same wire width. The result shows that for a large wire width, a ballistic p-FET delivers about one-half the ON-current of a ballistic n-FET. For a smaller wire width, however, the ON-current of the ballistic pFET approaches that of its n-type counterpart. To explain this observation, we plot the average carrier velocity (under high drain bias) *vs.* gate bias for the n-type/p-type SNWTs with *D*=1.36nm and *D*=5.15nm (Fig. 3 (b)). Interestingly, the carrier velocity for nFETs decreases while that for pFETs increases when the wire width is reduced. The origin of these trends involves different physics: 1) Due to the non-parabolicity of the four $\Delta$ valleys in the bulk Si conduction band, the projected $\Gamma$ valleys in the wires exhibit a larger transport effective mass at a smaller wire width (see the inset of Fig. 3 (b)), where stronger QC occurs. Since the electron thermal velocity is inversely proportional to the square root of the transport effective-mass,[13, 18] the average electron velocity in n-type SNWTs decreases as the wire width scales down. 2) For the valence band of the wires, our simulations show that the curvature of the valence band edge (hole effective-mass) is insensitive to the wire width. However, with increasing wire width, more and more higher subbands with larger transport hole effective-masses become populated (see Fig. 1 (b)), effectively lowering the average hole velocity in p-type SNWTs.

It is of great interest to compare SNWTs *vs.* planar MOSFETs. Previous studies[7, 8] show that the SNWT obtains a better gate control as well as a larger threshold voltage variation than the planar MOSFET due to its stronger QC. In this work, we compare the performance of SNWTs *vs.* planar MOSFETs in terms of carrier injection velocity. Figure 4 shows the average carrier velocities (under high drain bias) for the simulated SNWTs ($D$=1.36nm, 5.15nm) and the planar MOSFETs with comparable cross sections ($T_{Si}$=1.36nm, 5.15nm). (The E-k relations for the planar MOSFETs are calculated with the same tight binding approach as used in the nanowire calculation.) The normalized Fermi level, $\eta_F$, is defined as $\eta_F = (E_F - E_C)/k_B T$ for n-FETs and $\eta_F = (E_V - E_F)/k_B T$ for p-FETs, where $E_F$ is the Fermi level and $E_C$ ($E_V$) is the conduction (valence) band edge. The results show that for the same normalized Fermi level the p-SNWT (dashed line) displays a ~20% lower hole velocity as compared to the p-MOSFET (dashed line with open circles) at both $D(T_{Si})$=1.36nm and $D(T_{Si})$=5.15nm. For the n-type FETs, however, the SNWT (solid line) obtains a higher electron velocity than the planar MOSFET (solid line with closed circles) at $D(T_{Si})$=5.15nm while a lower one at $D(T_{Si})$=1.36nm. An explanation of this interesting observation requires an understanding of the role of QC on the *electron* velocity in a Si nanowire/thin-film.

Generally speaking, QC affects electron velocity in two opposite ways. First, QC lifts the 6-fold degeneracy of the Δ valleys in bulk Si so that the unprimed valleys (i.e., [010], [0$\bar{1}$0], [001] and [00$\bar{1}$] for a [100] oriented nanowire, or [001] and [00$\bar{1}$] for thin-films with a [001] confinement direction) with a relatively small transport effective-mass ($m_t = 0.19 m_e$ in bulk Si) display a smaller conduction band minimum and acquire a higher electron occupancy. As a result, the *average* electron velocity is increased by this valley-splitting effect caused by QC. Second, when the wire width or the film thickness is small, strong QC increases the transport effective mass (see the inset of Fig. 3 (b)) and consequently decreases the average electron velocity. From the results shown in Fig. 4, we conclude that when $D(T_{Si})$ is relatively large (e.g., 5.15nm), the valley-splitting effect dominates, and the wire obtains a larger electron velocity due to the additional QC in the width direction. When $D(T_{Si})$ is small (e.g., 1.36nm), however, the unprimed valleys are well separated from the primed valleys in both the wire and the thin-film, and the effective-mass-raising effect dominates. Thus, the wire displays a smaller electron velocity than the thin-film due to stronger QC.

In summary, by using an sp$^3$d$^5$s$^*$ tight binding model, we explored the energy dispersion relations of [100] oriented rectangular Si nanowires with a wire width up to 6.8nm. Based on these E-k relations, we calculated the ballistic currents for both n-type and p-type SNWTs with the use of a semi-numerical ballistic FET model. We found that for very small diameters ($D < 1.5$nm), ballistic p-SNWTs approach the performance of n-SNWTs. The carrier injection velocity for SNWTs was compared with those for planar MOSFETs, and we observed that p-SNWT displays a ~20% lower hole velocity than the p-type planar MOSFET (at both $D(T_{Si})$=1.36nm and $D(T_{Si})$=5.15nm). For nFETs, however, due to the effects of quantum confinement, the SNWT displays a higher electron velocity than the planar MOSFET when the wire width (or film thickness) is relatively large (e.g., $D(T_{Si})$=5.15nm) while a lower one when $D(T_{Si})$ is small (e.g., 1.36nm). In short, the bandstructure effects play an important role in nanowires with small diameters and should be seriously considered when evaluating the performance limits of SNWTs.

This work is supported by the Semiconductor Research Corporation (SRC), the Microelectronics Advanced Research Corporation (MARCO) focus center on Materials, Structures and Devices (MSD) and the National Science Foundation (NSF) Network for Computational Nanotechnology (NCN). The authors would like to thank Prof. Supriyo Datta at Purdue University for the useful discussions.

**References:**

[1]Y. Cui, Z. Zhong, D. Wang, W. U. Wang, C. M. Lieber, Nano Lett., 3(2), 149 (2003).

[2]D. D. D. Ma, C. S. Lee, F. C. K. Au, S. Y. Tong, S. T. Lee, Science, 299 (5614): 1874-1877 (2003).

[3]Y. Wu, Y. Cui, L. Huynh, C. J. Barrelet, D. C. Bell, C. M. Lieber, Nano Lett., 4(3), 433 (2004).

[4]L. Samuelson, M. T. Bjork, K. Deppert, M. Larssonb, B. J. Ohlssonc, N. Paneva, et al., Physica E, 21 (2-4), 560 (2004).

[5]H. Majima, Y. Saito and T. Hiramoto, Int. Electron Dev. Meet. Tech. Digest, 733 (2001).

[6]J. Wang, E. Polizzi and M. Lundstrom, J. App. Phys., 96 (4), 2192 (2004).

[7]J. Wang, E. Polizzi and M. Lundstrom, Int. Electron Dev. Meet. Tech. Digest, 695 (2003).

[8]J. Wang, P. Solomon and M. Lundstrom, IEEE Trans. Electron Dev., 51 (9), 1366 (2004).

[9]X. Zhao, C. Wei, L. Yang, M. Y. Chou, Phys. Rev. Lett., 92 (23), Art. No. 236805 (2004).

[10]Y. J. Ko, M. Shin, S. Lee and K. W. Park, J. App. Phys., 89 (1), 374 (2001).

[11]Y. Zheng, C. Rivas, R. Lake, K. Alam, T. B. Boykin and G. Klimeck, submitted (2004).

[12]T. B. Boykin, G. Klimeck and F. Oyafuso, Phys. Rev. B, 69, 115201/1-10 (2004).

[13]A. Rahman, J. Guo, S. Datta and M. Lundstrom, IEEE Trans. Electron Dev., 50 (9), 1853 (2003).

[14]S. Lee, F. Oyafuso, P. von Allmen and G. Klimeck, Phys. Rev. B, 69, 045316 (2004).

[15]T. B. Boykin, G. Klimeck, M. A. Eriksson, M. Friesen, S. N. Coppersmith, P. Von Allmen, et al., App. Phys. Lett., 84 (1), 115 (2004). A. Rahman, G. Klimeck, M. Lundstrom, N. Vagidov, and T. B. Boykin, submitted (2004).

[16]S. Luryi, Appl. Phys. Lett, 52, 501 (1988).

[17]J. Wang and M. Lundstrom, IEEE Trans. Electron Dev., 50 (7), 1604, (2003).

[18]A. Rahman, G. Klimeck and M. Lundstrom, to appear in Int. Electron Dev. Meet. Tech. Digest (2004).

[19]http://nanohub.org

# Figure Captions:

FIG. 1  The energy dispersion relations for the simulated Si nanowire structures with (a) $D$=1.36nm and (b) $D$=5.15nm. The inset shows a schematic diagram of the nanowire cross section ($T_{Si}=W_{Si}=D$). For the thinner wire ($D$=1.36nm), strong band splitting is observed at the $\Gamma$ point in the conduction band.

FIG. 2  The conduction (upper) and valence (lower) band edges (solid with circles) at $\Gamma$ point *vs.* $D$. The dashed line in the upper plot is for the conduction band edge calculated by the effective-mass (EM) approach with a simple "particle in a box" model.

FIG. 3  (a) The *I-V* curves for the n-type/p-type SNWT with $D$=1.36nm and the ratio of the pFET ON-current to the nFET's *vs.* $D$ (inset). The oxide thickness is assumed to be 1nm and the temperature is 300K. (b) The average carrier velocities (under high drain bias) *vs.* gate bias for the n-type/p-type SNWTs with $D$=1.36nm and $D$=5.15nm. The inset shows the dependence of the transport effective mass (in the conduction band) at $\Gamma$ point on $D$. The carrier velocities for the nFET and the pFET show different trends with decreasing $D$ due to different physics.

FIG. 4  Average carrier injection velocity (under high drain bias) *vs.* normalized Fermi level, $\eta_F$, for the simulated SNWTs and planar MOSFETs. $D$ ($T_{Si}$) is equal to 1.36nm (left) and 5.15nm (right) for the SNWTs (planar MOSFETs).

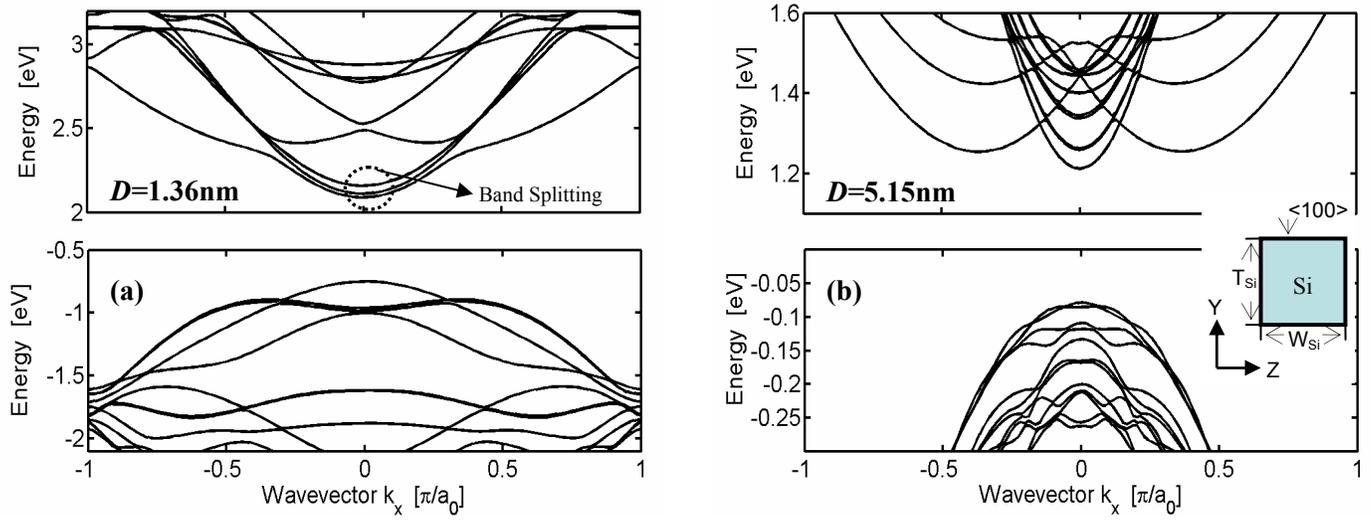

FIG. 1  The energy dispersion relations for the simulated Si nanowire structures with (a) $D$=1.36nm and (b) $D$=5.15nm. The inset shows a schematic diagram of the nanowire cross section ($T_{Si}=W_{Si}=D$). For the thinner wire ($D$=1.36nm), strong band splitting is observed at the $\Gamma$ point in the conduction band.

FIG. 2  WANG, RAHMAN, GHOSH, KLIMECK and LUNDSTROM

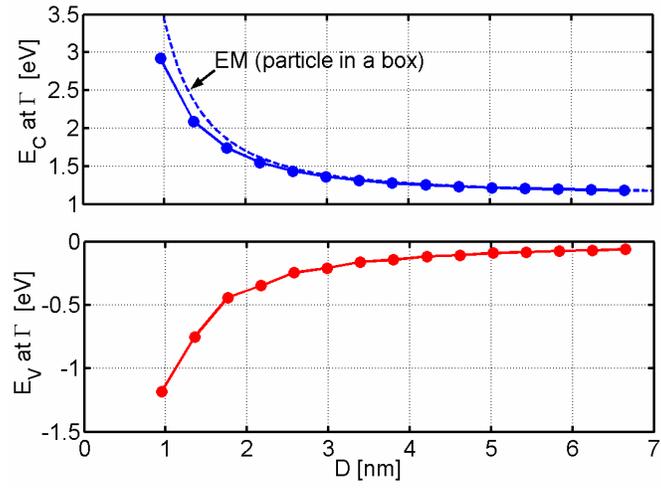

FIG. 2  The conduction (upper) and valence (lower) band edges (solid with circles) at Γ point *vs. D*. The dashed line in the upper plot is for the conduction band edge calculated by the effective-mass (EM) approach with a simple "particle in a box" model.

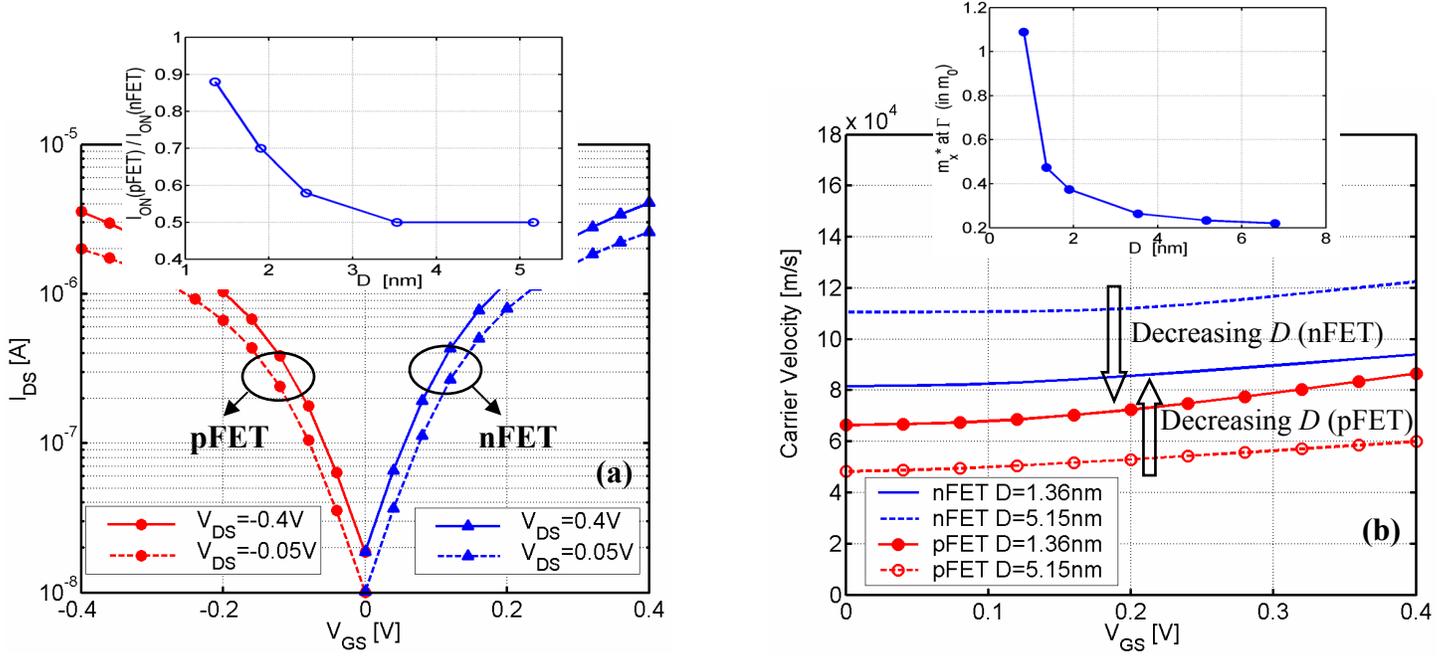

FIG. 3  (a) The *I-V* curves for the n-type/p-type SNWT with *D*=1.36nm and the ratio of the pFET ON-current to the nFET's *vs. D* (inset). The oxide thickness is assumed to be 1nm and the temperature is 300K. (b) The average carrier velocities (under high drain bias) *vs.* gate bias for the n-type/p-type SNWTs with *D*=1.36nm and *D*=5.15nm. The inset shows the dependence of the transport effective mass (in the conduction band) at Γ point on *D*. The carrier velocities for the nFET and the pFET show different trends with decreasing *D* due to different physics.

FIG. 4 WANG, RAHMAN, GHOSH, KLIMECK and LUNDSTROM

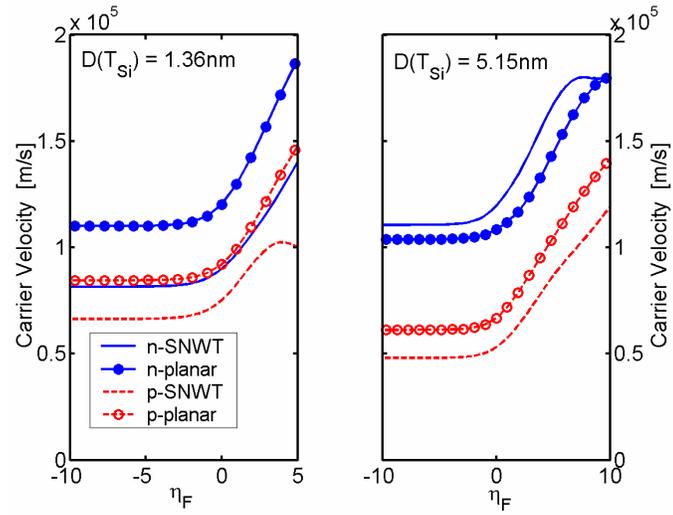

FIG. 4 Average carrier injection velocity (under high drain bias) *vs.* normalized Fermi level, $\eta_F$, for the simulated SNWTs and planar MOSFETs. $D$ ($T_{Si}$) is equal to 1.36nm (left) and 5.15nm (right) for the SNWTs (planar MOSFETs).